# The *Swift* X-Ray Telescope: Status and Performance

David N. Burrows[*,a], J. A. Kennea[a], A. F. Abbey[b], A. Beardmore[b], S. Campana[c],
M. Capalbi[d], G. Chincarini[c], G. Cusumano[e], P. A. Evans[b], J. E. Hill[f,g], P. Giommi[d], M. Goad[b],
O. Godet[b], A. Moretti[c], D. C. Morris[a], J. P. Osborne[b], C. Pagani[a], K. L. Page[b], M. Perri[d],
J. Racusin[a], P. Romano[c], R. L. C. Starling[b], G. Tagliaferri[c], F. Tamburelli[d], L. G. Tyler[b],
and R. Willingale[b]

[a]Pennsylvania State University, 525 Davey Lab, University Park, PA 16802, USA
[b]University of Leicester, Leicester LE1 7RH, UK
[c]Osservatorio Astronomico di Brera, Via Brera 28, 20121 Milano, Italy
[d]ASI Science Data Center, Frascati, Italy
[e]INAF - Istituto di Astrofisica Spaziale e Fisica Cosmica, Palermo, Italy
[f]Universities Space Research Association, Columbia, MD 21044, USA
[g]CRESST and NASA Goddard Space Flight Center, Greenbelt, MD 20771, USA

**ABSTRACT**

We present science highlights and performance from the Swift X-ray Telescope (XRT), which was launched on November 20, 2004. The XRT covers the 0.2-10 keV band, and spends most of its time observing gamma-ray burst (GRB) afterglows, though it has also performed observations of many other objects. By mid-August 2007, the XRT had observed over 220 GRB afterglows, detecting about 96% of them. The XRT positions enable followup ground-based optical observations, with roughly 60% of the afterglows detected at optical or near IR wavelengths. Redshifts are measured for 33% of X-ray afterglows. Science highlights include the discovery of flaring behavior at quite late times, with implications for GRB central engines; localization of short GRBs, leading to observational support for compact merger progenitors for this class of bursts; a mysterious plateau phase to GRB afterglows; as well as many other interesting observations such as X-ray emission from comets, novae, galactic transients, and other objects.

Keywords: gamma-ray burst, X-ray telescope, Swift, X-ray instrumentation, X-ray CCD detector, X-ray mirrors

## 1 INTRODUCTION

The *Swift* Gamma Ray Burst Explorer[1] was launched on November 20, 2004. It carries three instruments: a *Burst Alert Telescope (BAT* [2]*)*, which identifies gamma-ray bursts (GRBs) and determines their location on the sky to within a few arcminutes; an *Ultraviolet/Optical Telescope (UVOT* [3]*)* with sensitivity down to 24$^{th}$ magnitude and 0.3 arcsecond positions; and an *X-ray Telescope* (*XRT* [4]). The three instruments combine to make a powerful multi-wavelength observatory with the capability of rapid position determinations of GRBs to arcsecond accuracy within 1-2 minutes of their discovery, and the ability to measure both light curves and redshifts of the bursts and afterglows.

The *Swift XRT* is a sensitive, flexible, autonomous X-ray imaging spectrometer designed to measure fluxes, spectra, and light curves of GRBs and afterglows over a wide dynamic range of more than 7 orders of magnitude in flux.

It utilizes a Wolter I mirror[5] and an e2v CCD-22 detector[6,7] to provide a sensitive broad-band (0.2-10 keV) X-ray imager with effective area of 120 cm$^2$ at 1.5 keV[8], field of view of 23.6 x 23.6 arcminutes, and angular resolution of 18 arcseconds (HEW)[9]. The instrument is designed to provide automated source detection and position reporting within 5 seconds of target acquisition[10]. The *XRT* usually operates in an auto-exposure mode, adjusting the CCD readout mode automatically to optimize the science return for each frame as the source fades.

## 2 XRT PERFORMANCE

### 2.1 POSITION DETERMINATION

The prime science requirement for the XRT is to produce accurate positions of GRB afterglows. Here we describe several types of position determinations made by XRT on different timescales, and compare their performance.

---

[*] burrows@astro.psu.edu; phone +1 (814) 865-7707; fax +1 (814) 865-9100; http://www.swift.psu.edu/xrt

These are discussed in the order in which they become available after a GRB is detected by *Swift*.

### 2.1.1 On-board Centroids

GRB afterglows are often quite bright at early times, far brighter than typical background sources (see Fig. 1), and the rapid slewing capability of the *Swift* observatory means that XRT count rates are often between 10 and 100 counts per second when the instrument is first pointed at a new burst (typically within 2 minutes after the burst occurs). The XRT takes advantage of this by taking a 0.1 s image and a 2.5 s image immediately after the slew to the GRB ends, and attempting to find a bright X-ray source in either image. If a source is found, its centroid position is reported to the ground via the TDRSS satellite network for distribution to the astronomical community[11]. While rapid, the accuracy of these centroids is limited by the low photon statistics typically available and by the fact that the satellite is still drifting slowly, as well as by detector artifacts (such as hot pixels and columns) and cosmic ray events[43]. Nevertheless, typical accuracy is of order 5-6 arcseconds at very early times. Figure 2 shows the offsets in sky position between XRT on-board centroid positions and optical positions for all GRB afterglows that have both XRT centroid and optical positions. Figure 3 shows these offsets as a function of GRB number (roughly proportional to time).

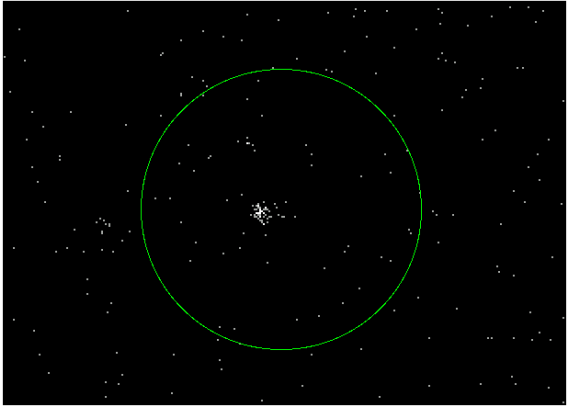

Figure 1: Typical XRT afterglow. Each dot represents an X-ray photon. The circle is the BAT error circle for the burst. XRT afterglows are typically the brightest object inside the BAT error circle.

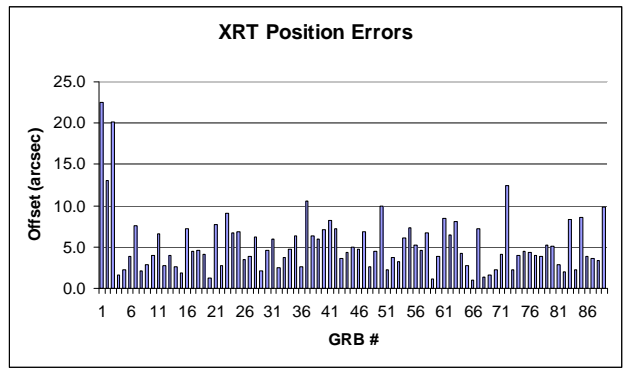

Figure 3: Errors between on-board centroids and optical positions of GRB afterglows, as a function of GRB number. The first few GRBs, with high errors, were before calibration of the instrument boresight. After that point, typical errors are less than 5 arcseconds.

Excluding the first few GRBs, which occurred before completion of boresight calibration, the mean offsets between X-ray and optical positions is -1.1 arcseconds in Right Ascension (RA) and -0.8 arcseconds in Declination. The mean (median) offset between X-ray and optical positions using this method is 4.8 (4.4) arcseconds.

### 2.1.2 Prompt Ground-based Positions

Since May 1, 2006, the XRT has been using a new on-board software feature that transmits single-pixel X-ray events to the ground via TDRSS during the first orbit's worth of observations of a new GRB. These data, referred to as SPER data, allow a relatively rapid calculation of the GRB afterglow position using ground-based software that is more accurate than the on-board centroiding software. These improved SPER positions are generally available within 5-10 minutes after the burst, except in cases where the count rate is so high that no Photon-Counting mode data[12] are acquired during the first orbit. Beginning in August 2007, the XRT team distributes

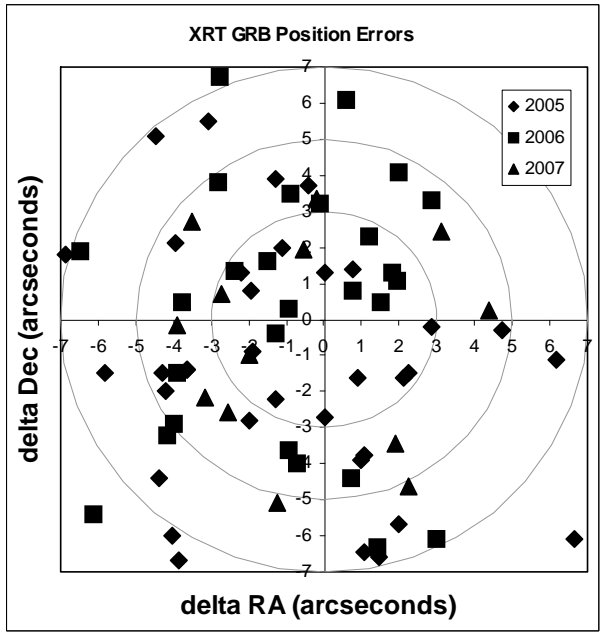

Figure 2: Position errors between on-board centroids and optical positions for GRB afterglows. Circles are 3", 5", and 7".

XRT Position Notices to the astronomical community when the first SPER position becomes available for each burst. (We do not continue to distribute notices each time the position is improved with subsequent SPER data.)

The accuracy of the SPER messages is illustrated in Figure 4, which shows offsets between XRT and optical positions for GRBs with both SPER and optical positions. The mean offsets are 0.2 arcseconds in RA and 0.1 arcseconds in Declination. The X-ray/optical offsets are shown as a function of GRB number in Figure 5. The mean (median) offset is 3.2 (2.6) arcseconds, significantly better than the on-board positions. The improvement is partly the result of better photon statistics, partly because the spacecraft has stopped drifting when these data are produced, and partly because of an improved PSF fitting routine discussed in section §2.1.4.

### 2.1.3 Ground-based positions

On a typical time-scale of 2-3 hours after the burst a ground-processed position can be determined using the full data set downloaded to the Malindi ground station. The techniques used in this position determination have been described previously[13]. The resulting position accuracy is shown in Figure 6. Most of the X-ray positions are within 5 arcseconds of the optical positions. The mean offsets are -0.07 arcseconds in RA and -0.2 arcseconds in Declination. The offsets are shown in Figure 7; the mean (median) offset is 2.6 (2.5) arcseconds.

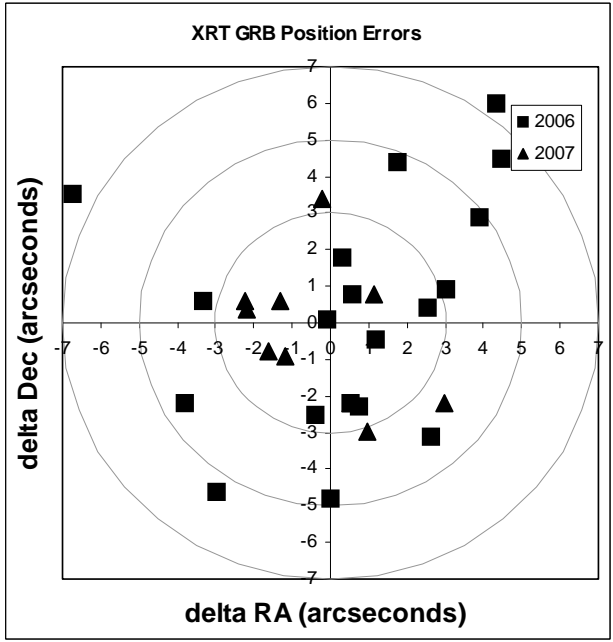

Figure 4: Position errors for SPER positions for GRBs with both SPER X-ray positions and optical positions.

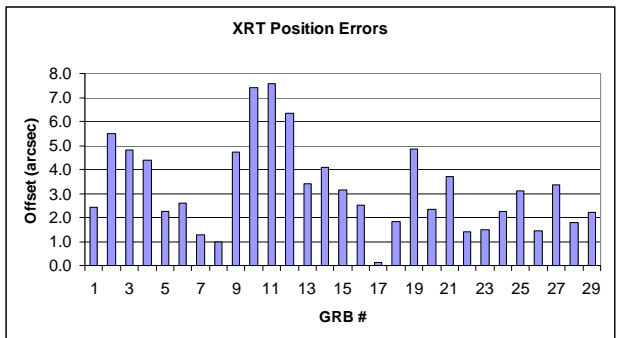

Figure 5: X-ray and optical offsets for SPER positions.

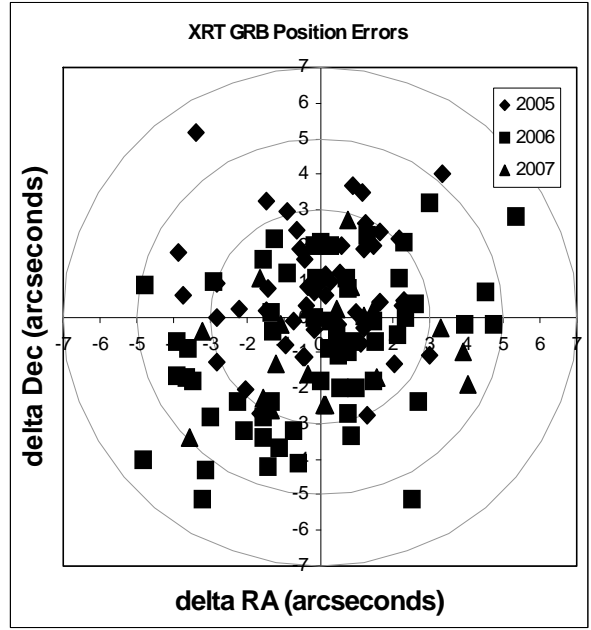

Figure 6: Position errors for refined ground-calculated positions for GRB with optical positions.

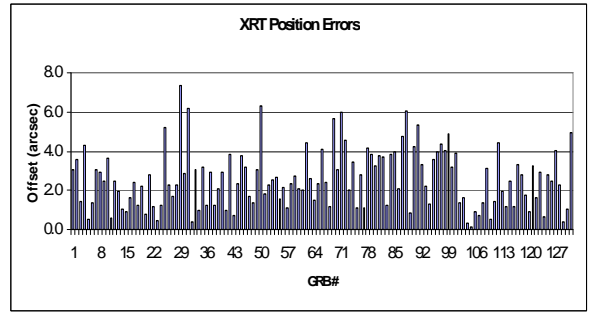

Figure 7: X-ray and optical offsets for ground-calculated refined GRB positions. The mean position error is 2.6 arcseconds.

### 2.1.4 Astrometric positions

The uncertainties in the techniques listed above are dominated by two components: photon statistics, and the uncertainty in the star tracker solutions, which imposes a lower limit of about 3 arcseconds on the XRT position accuracy. We have recently developed a technique[14] that eliminates the latter (star tracker) component by using the UVOT instrument as a "super star tracker", taking advantage of the precise registration achievable between UVOT images and astrometric star catalogs, and of the highly stable *Swift* Optical Bench, which maintains the relative alignment between the XRT and UVOT boresights to better than an arcsecond in flight. We achieve additional improvements by using a fitting routine to determine XRT source positions in detector coordinates that accounts for the instrument PSF and that correctly and accurately treats the effects of bad pixels and columns. The resulting position errors are shown in Figure 8. The mean offsets are 0.01 arcseconds in RA and -0.08 arcseconds in Declination. The position offsets are shown in Figure 9; the mean (median) offset is 1.9 (1.4) arcseconds for these "UVOT-enhanced" positions. We note that at these small offsets, errors in the optical positions become significant. We have used optical positions given in GCN Circulars, most of which do not identify the astrometric system used to derive them, and many of which do not specify their as-trometric uncertainties. In these figures we have simply plotted the difference between the best X-ray and optical positions. Testing of this technique using bursts with UVOT afterglows shows that 90% of the XRT positions are within 2 arcseconds of the UVOT position[14].

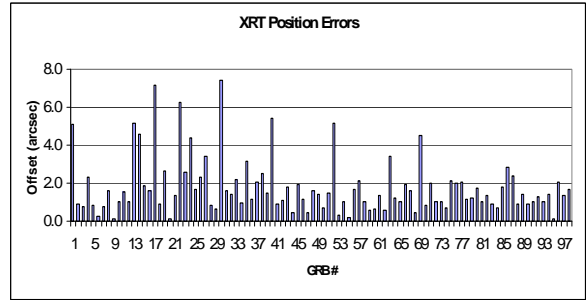

Figure 9: Position offsets between UVOT-enhanced X-ray positions and optical positions vs GRB number. We note that the GRB numbering is different for each of these plots, and simply represents the ordinal number of the GRB within the particular subsample being plotted.

The procedure used to produce these "UVOT-enhanced" positions has been described in detail elsewhere[14]. These positions are available for about 60% of *Swift* bursts (availability requires that the XRT obtain PC mode data simultaneously with UVOT V band images). For new bursts, these improved positions are generally available within a few hours after the burst (i.e., on a similar time-scale to the standard refined ground-processed positions), and are distributed automatically to the community when they become available. The position accuracy achieved with this technique is comparable to that achieved through astrometric positions based on matching serendipitous sources in deep XRT exposures to astrometric catalogs, but on time scales of hours rather than days or weeks.

### 2.2 SPECTROSCOPY

Prior to launch, we expected that some of the most exciting results from the *Swift* XRT would be related to measurements of emission lines in X-ray spectra. Previous work had found evidence at later times for spectral lines from Fe and other heavy elements[15,16], and it was thought that with much earlier observations, when the afterglow was orders of magnitude brighter, the XRT might be able to confirm the previous rather marginal detections, and even to use line detections to measure redshifts and place interesting constraints on the emission regions.

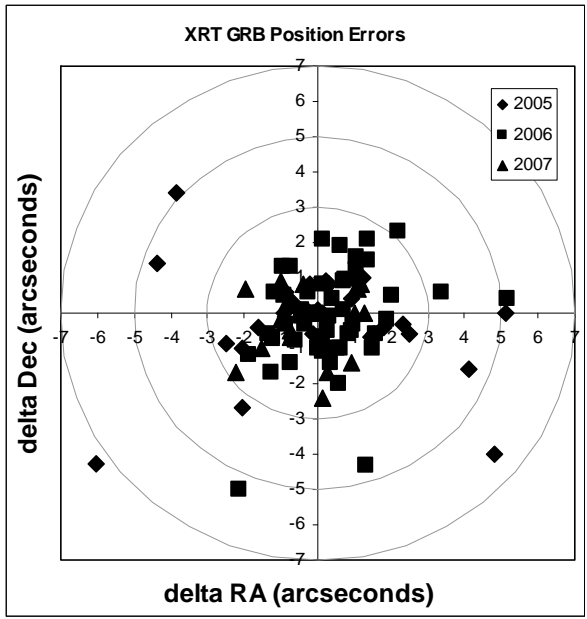

Figure 8: Position errors for XRT astrometrically-corrected (UVOT-enhanced) positions[14] for bursts with optical counterparts. Note that errors in the optical positions are no longer negligible at these offset values, and may contribute significantly.

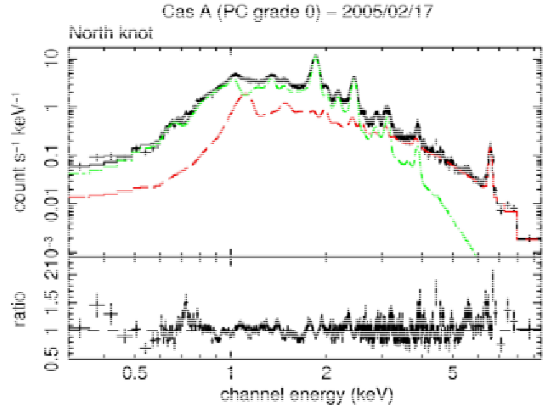

Figure 10: *Swift* XRT spectrum of the northern knot of the Cas A supernova remnant. The data are fit by an absorbed 2 component non-equilibrium ionization spectral model that fits the data from the XMM MOS1 spectrum of this knot, scaled by a constant factor to account for uncertainties in the effective areas of the two instruments. The spectrum is dominated by strong lines of Si, S, Ar, Ca, and Fe.

This has not turned out to be the case. The spectral resolution of the XRT is illustrated in Figure 10, which shows a spectrum of the northern knot of the Cas A supernova remnant taken in February 2005 during our on-orbit calibration. The spectrum is dominated by strong lines of hydrogen-like and helium-like Si, S, Ar, Ca, and Fe. By contrast, we show a typical GRB spectrum in Figure 11. This spectrum is fit by a simple absorbed power law, with no hint of any spectral lines. We occasionally find other spectral components: flares often exhibit a more complex spectrum described by a Band function or cutoff power law, and a few bursts have soft excesses that can be fit with a blackbody component, but we have seen no spectral lines in any GRB afterglow[17].

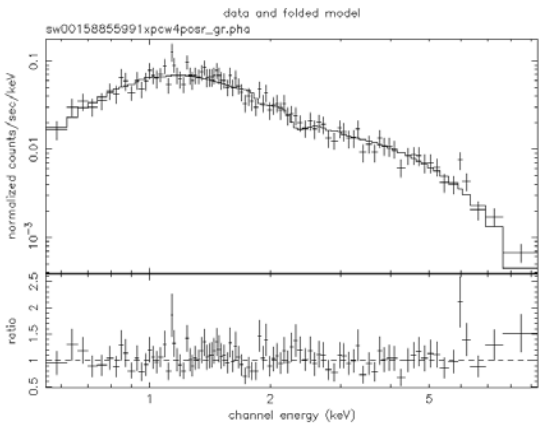

Figure 11: XRT spectrum of GRB 051008. The best fit is to a simple absorbed power law.

### 2.3 TIMING

The third important capability of the XRT is high resolution timing. Although many interesting results have come from XRT light curves of GRB afterglows, no interesting periodicities or other high frequency temporal phenomena have been identified to date.

## 3 SCIENCE HIGHLIGHTS

Scientific implications of the *Swift* XRT data have been published elsewhere in journals and conference proceedings. Here we will simply highlight a few of the key results.

### 3.1 XRT AFTERGLOW STATISTICS

As of mid-August 2007, the XRT has observed 231 GRBs detected by the *Swift* BAT instrument (as well as several dozen detected by *HETE-II*, *INTEGRAL*, and *AGILE*). The XRT has detected 96% of the BAT-discovered bursts (221 detected X-ray counterparts). This represents more than 4 times the total sample of X-ray afterglows of GRBs observed before the launch of *Swift*. The sample of X-ray afterglows with prompt observations includes 181 bursts (most of the remainder had observations delayed by Earth occultations). The mean (median) time on target for prompt XRT observations of BAT-detected bursts was 103 (95) seconds after the burst trigger.

GRBs come in (at least) two types, generally called long and short GRBs[18]. Long GRBs are thought to originate in the collapse of massive stars. The XRT has detected 98% of the 202 BAT-discovered long GRBs it has observed to date. The XRT detection rate of short GRBs is considerably lower, with 19 out of 24 short GRBs discovered by the BAT being detected by the XRT. The short GRBs tend to have much weaker afterglows, and many of them fade quickly, with only a few dozen photons detected.

Because of the high detection rate for X-ray afterglows and the small error circles provided by the XRT, X-ray detections often facilitate the discovery of optical afterglows. The redshift distribution of long *Swift* GRBs is shown in Figure 12. Reliable redshifts have been determined for about 1/3 of the bursts with XRT detections. The mean redshift through mid-August 2007 is 2.4, which is about twice the mean redshift for pre-*Swift* GRBs. Five *Swift* GRBs have redshifts greater than 5.0: GRB 050814 (5.3), GRB 050904 (6.3), GRB 060116 (6.6), GRB 060522 (5.1), and GRB 060927 (5.6).

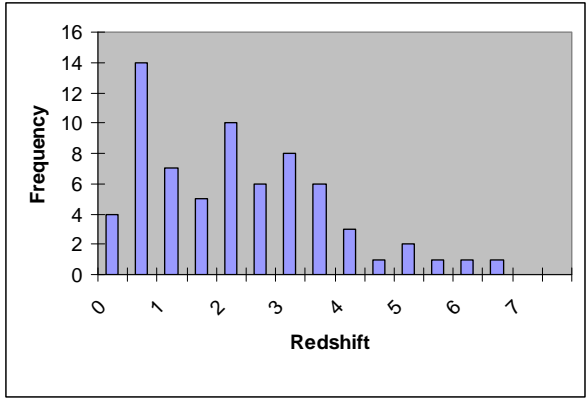

Figure 12: Redshift distribution of long *Swift* GRBs through mid-August 2007. The mean (median) redshift is 2.39 (2.35). The highest redshift is a photometric redshift of 6.6 for GRB 060116.

### 3.2 X-RAY FLARES

The typical GRB afterglow is expected to fade as a power law in time, and many GRB X-ray afterglows do decay as a series of power laws of differing decay index, but the XRT data show that many afterglows exhibit much more complex behavior. X-ray flares have been found in a large fraction of *Swift* GRB afterglows[19-25]. The flares are characterized by large, rapid flux variations. The largest flare seen to date had fluence equal to that of the prompt γ-ray emission[21]. The probability of flare occurrence decreases linearly with time since the burst. The most likely explanation for the flares is that the "central engine" of the GRB continues to be intermittently active long after the burst of γ-rays ends. Indeed, the central engine activity must continue for times up to hours after the burst (Fig 13).

### 3.3 RAPID DECAY PHASE

An unexpected discovery is that many X-ray light curves begin with an extremely steep decay, of order $t^{-3}$ or steeper, during which the flux drops by several orders of magnitude for some tens of minutes (Figure 14)[27-30]. This phase has been explained as the result of the end of the prompt emission, and although explained theoretically before the launch of *Swift*[31], had not previously been seen.

### 3.4 THE PLATEAU PHASE

Another unexpected feature of the X-ray light curves is a plateau phase, in which the decaying light curve flattens to a very low decay index (nearly constant in time) for a period of several hours. An example is the light curve of GRB 060729[32] (Figure 14). This plateau has been interpreted as evidence for quasi-continuous injection of energy into the external shock with the ambient medium[28-29], but the unusual plateau phase of GRB 070110[33], which terminates very abruptly, cannot be explained by this mechanism, suggesting that at least some plateaus are related to quasi-continuous late-time central engine activity.

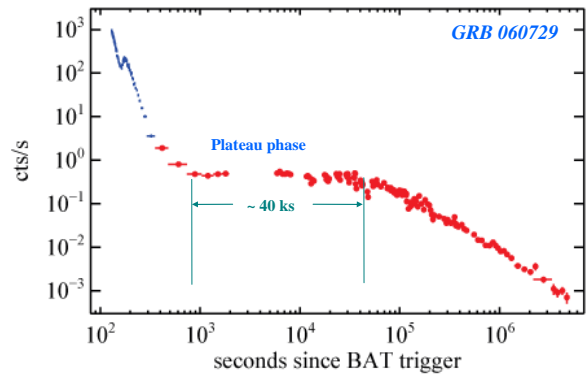

Figure 14: The extraordinarily long afterglow of GRB 060729, the longest X-ray afterglow observed to date by the *Swift* XRT[32].

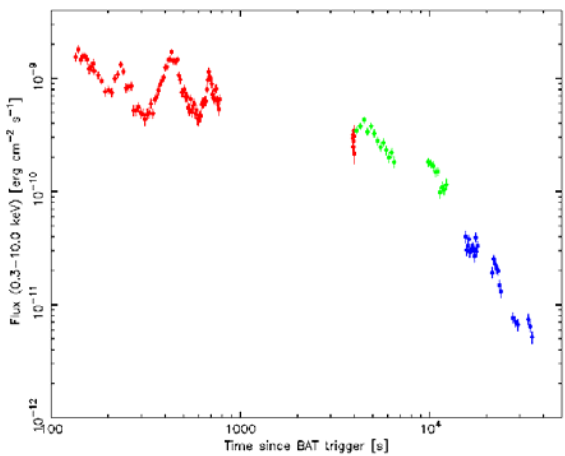

Figure 13: The flaring X-ray afterglow of GRB 050730[26]. Flaring continues out to at least 30,000 s after the burst (in the observer frame).

## 3.5 SHORT GRB LOCALIZATIONS

Short GRBs had eluded localization until May 9, 2005, when the *Swift* BAT detected GRB 050509B and the XRT localized it to within 10 arcseconds of a giant elliptical galaxy in a cluster[34]. This was remarkable in several regards. No long GRB had ever been localized to an elliptical galaxy. In fact, no long GRB had ever been localized to any galaxy in a cluster. Long GRBs are typically associated with small irregular galaxies and are localized to regions of active star formation, consistent with their deduced origin in the demise of massive stars, which are born in star forming regions and which evolve rapidly and die young. Elliptical galaxies, on the other hand, contain only very old stars, and have no current star formation. This clearly points to a different origin for short GRBs than for long ones, and supports theories that short GRBs may originate in the merger of two neutron stars or a neutron star and a black hole, rather than in the collapse of a massive star[35,36]. Subsequent short GRB localizations generally support this picture[37], although the situation is complicated by a lack of clear host galaxies for many short bursts, and by the lack of any short bursts for which the distance has been measured directly by red-shifted absorption lines in the light of the optical afterglow (instead, short GRB distances have been measured by spectral lines from the host galaxy, which is often ambiguous).

## 3.6 OTHER SCIENCE

When *Swift* is not observing GRB afterglows, it performs a wide variety of other science observations, many of which rely on the XRT for their primary results. As part of its active Target of Opportunity (ToO) program, the XRT has provided precise positions of both transient and steady sources discovered by the *Swift* BAT instrument, by the *INTEGRAL* satellite, and by the *RXTE* satellite. We have observed objects ranging from comets[38] to novae[39] to supernovae[40] and active galactic nuclei[41]. The success of these observations clearly demonstrates the utility and importance of an X-ray telescope with rapid response capability (the XRT has executed a ToO observation in as little as 40 minutes after the event was discovered, and can often respond within an hour or two for high priority ToOs).

## 4 CONCLUSIONS

The Swift XRT has been remarkably successful in its first two years of operation. The instrument is healthy and is working well, in spite of the loss of the active cooling system shortly after launch, and the effects of a micrometeoroid impact on the detector in May 2005[42]. Swift has no expendables and we expect it to continue operating for many years, expanding our knowledge of GRB physics and transient phenomena in our Galaxy and in the Universe.

## 5 ACKNOWLEDGEMENTS

This work is supported at Penn State by NASA contract NAS5-00136; at the University of Leicester by the Particle Physics and Astronomy Research Council; and at OAB, Palermo, and the ASDC by funding from ASI. We gratefully acknowledge the contributions of dozens of members of the *XRT* team at PSU, UL, OAB, GSFC, and our subcontractors, who helped make this instrument possible.

In particular, we would like to acknowledge the contributions of our friend and colleague, Francesca Tamburelli, who passed away earlier this year. Francesca was part of the *Beppo-SAX* Science Data Center and the ASI Science Data Center from their inception. She was the main architect behind the *Swift* and *AGILE* data systems. We have lost a bright mind and a dear friend.